\documentclass[english,notitlepage,showpacs,preprintnumbers,amsmath,amssymb,aps,twocolumn]{revtex4-1}
\usepackage[T1]{fontenc}
\usepackage[latin9]{inputenc}
\usepackage{amstext}
\PassOptionsToPackage{normalem}{ulem}
\usepackage{ulem}

\makeatletter

\providecommand{\tabularnewline}{\\}

\@ifundefined{textcolor}{}
{%
 \definecolor{BLACK}{gray}{0}
 \definecolor{WHITE}{gray}{1}
 \definecolor{RED}{rgb}{1,0,0}
 \definecolor{GREEN}{rgb}{0,1,0}
 \definecolor{BLUE}{rgb}{0,0,1}
 \definecolor{CYAN}{cmyk}{1,0,0,0}
 \definecolor{MAGENTA}{cmyk}{0,1,0,0}
 \definecolor{YELLOW}{cmyk}{0,0,1,0}
}

\usepackage{amsfonts}
\usepackage{palatino}
\usepackage{MnSymbol}
\usepackage{bbm}
\usepackage{hyperref}
\usepackage{accents} 

\usepackage{algorithm}
\usepackage{algpseudocode}
\usepackage{forest}

\algnewcommand\algorithmicappend{\textbf{append}}
\algnewcommand\Append{\algorithmicappend}
\algnewcommand\algorithmicTo{\textbf{to}}
\algnewcommand\To{\algorithmicTo}
\algrenewcommand\algorithmicrequire{\textbf{Input:}}
\algrenewcommand\algorithmicensure{\textbf{Output:}}

\makeatother

\usepackage{babel}
\begin{document}

\title{Reducing multi-qubit interactions in adiabatic quantum computation.
Part 2: The ``split-reduc'' method and its application to quantum
determination of Ramsey numbers}

\author{Emile Okada$^{1,\,}$}

\email{eto25@cam.ac.uk}

\affiliation{$^{1}$Department of Mathematics, Cambridge University, CB2 3AP,
Cambridge, UK. }

\author{Richard Tanburn$^{2,\,}$}

\email{richard.tanburn@hertford.ox.ac.uk}

\affiliation{$^{2}$Mathematical Institute, Oxford University, OX2 6GG, Oxford,
UK. }

\author{Nikesh S. Dattani$^{3,4,\,}$}

\email{nike.dattani@gmail.com}

\affiliation{$^{3}$School of Materials Science and Engineering, Nanyang Technological
University, 639798, Singapore,}

\affiliation{$^{4}$Fukui Institute for Fundamental Chemistry, 606-8103, Kyoto,
Japan}
\begin{abstract}
Quantum annealing has recently been used to determine the Ramsey numbers
$R(m,2)$ for $4\le m\le8$ and $R(3,3)$ {[}Bian \emph{et al.} (2013)
PRL \textbf{111}, 130505{]}. This was greatly celebrated as the largest
experimental implementation of an adiabatic evolution algorithm to
that date. However, in that computation, more than 66\% of the qubits
used were auxiliary qubits, so the sizes of the Ramsey number Hamiltonians
used were tremendously smaller than the full 128-qubit capacity of
the device used. The reason these auxiliary qubits were needed was
because the best quantum annealing devices at the time (and still
now) cannot implement multi-qubit interactions beyond 2-qubit interactions,
and they are also limited in their capacity for 2-qubit interactions.
We present a method which allows the full qubit capacity of a quantum
annealing device to be used, by reducing multi-qubit and 2-qubit interactions.
With our method, the device used in the 2013 Ramsey number quantum
computation could have determined $R(16,2)$ and $R(4,3)$ with under
10 minutes of runtime. 
\end{abstract}
\maketitle

\section{Introduction}

The capacities and limits for adiabatic quantum computers (AQCs) to
outperform classical computers, and to speed-up the solution to discrete
optimization problems has recently been discussed in \cite{Tanburn2015a}.
As discussed in \cite{Tanburn2015a}, the quantum annealing devices
with the largest qubit capacities tend only to allow up to at most
2-qubit interactions, and are even limited in the 2-qubit interactions
allowed. Similarly, even when solving a discrete optimization problem
on a classical computer, high-order terms rapidly make the problem
more difficult. If only up to linear terms (1 qubit terms) are present
in the Hamiltonian (objective function), then finding the solution
to the problem is trivial, but if quadratic terms (2-qubit terms)
are allowed the problem becomes NP complete. 

Nevertheless, an enormous body of work has been done on efficient
algorithms for quadratic unconstrained Boolean optimization (QUBO)
problems, and it is known that if all coefficients of quadratic terms
are negative, the solution can be found in polynomial time \cite{Nemhauser1981,Orlin2007,Jegelka2011,Boros2002}.
When cubic (3-qubit) terms and beyond are present, another leap in
difficulty arises, and most of the effort is typically spent on quadratizing
such objective functions (Hamiltonians). Most quadratization techniques
work by adding auxiliary variables (qubits), and while algorithms
for finding solutions to discrete optimization problems often scale
exponentially with the number of variables, it is \emph{still} often
desirable to remove cubic terms and higher at the expense of adding
more variables. 

However, quantum annealing devices to date are very limited in qubit
capacity (the largest device reported to date having only about 2
kiloqubits or 258 qubytes). Therefore, adding auxiliary qubits is
usually not an option if any benefit over traditional computation
methods is desired for any relevant problem. In Part 1 \cite{Tanburn2015a}
we demonstrated a method called ``deduc-reduc'' which reduces multi-qubit
interactions without adding auxiliary qubits, and for the integer
factorization problem, managed to eliminate thousands of 4-qubit and
3-qubit interactions with just a few seconds of CPU time. A drawback
of this method is that some deduction must be made which relates the
variables of the discrete optimization problem (an example of such
a deduction could be $x_{1}+x_{2}=1$). Such deductions arise naturally
for the problem of integer factorization, but there is no reason to
believe that such deductions can be made for an arbitrary discrete
optimization problem. 

In this paper we present a method for reducing multi-qubit interactions
without adding auxiliary qubits \emph{and }without the need for any
deductions, but it increases the number of objective functions that
need to be minimized to find the solution to the original objective
function, and adding auxiliary qubits improves the method. We call
this method ``split-reduc'' since it iteratively \emph{\uline{splits}}
the Hamiltonian into separate Hamiltonians in order to \emph{\uline{reduce}}\emph{
}multi-qubit terms. We give very conservative lower and upper bounds
on the number of new objective functions created, and we showcase
split-reduc on the Hamiltonian used in the determination of Ramsey
numbers using quantum annealing, as in \cite{Bian2013}.

\section{A quick example\label{sec:A-quick-example}}

Let us demonstrate the method with the simple objective function $H=1+x_{1}x_{2}x_{5}+x_{1}x_{6}x_{7}x_{8}+x_{3}x_{4}x_{8}-x_{1}x_{3}x_{4}$
and an adiabatic quantum computer (AQC) that only has 8 qubits and
only allows up to at most 2-qubit interactions. Due to the restriction
on the number of qubits we cannot reduce the qubic terms to quadratic
terms by introducing auxiliary variables. The simple, but effective
idea is then to ``split'' the objective function into two by setting
a variable to its two possible values (0 or 1). In this case $x_{1}$
is the obvious choice to split over since it is present in the most
terms and contributes to the quartic term. Setting $x_{1}$ to 0 results
in the objective function $H_{0}=1+x_{3}x_{4}x_{8}$ and setting $x_{1}$
to 1 results in $H_{1}=1+x_{2}x_{5}+x_{6}x_{7}x_{8}+x_{3}x_{4}x_{8}-x_{3}x_{4}$. 

$H_{0}$ still contains cubic terms so we have the choice to split
$H_{0}$ further. However at this point we have 5 unused qubits so
we could also stop here by using one of them as an auxiliary variable
to quadratize the objective function. $H_{1}$ on the other hand is
however still a bit too complicated for our quantum computer to handle.
It contains cubic terms and requires 7 qubits out of the 8 qubit capacity
of the AQC, so we split again, this time over $x_{8}$. We get the
objective functions $H_{10}=1+x_{2}x_{5}+x_{6}x_{7}$ and $H_{11}=1+x_{2}x_{5}+x_{3}x_{4}$
. Both only contain quadratic terms so we have succeeded in turning
our Hamiltonian into 3 separate Hamiltonians that can each be implemented
on the AQC. In general, we can reduce the number of splits necessary,
by combining this approach with established methods for quadratization
techniques that introduce auxiliary variables.

\section{The method}

We now demonstrate the method in full generality. We first define
two cost functions:
\begin{enumerate}
\item $C(H)$ tells us whether or not we need to split the Hamiltonian any
further, and
\item \textbf{$C_{H}(x_{i})$ }tells us which variable to choose for the
splitting at each step, by assigning a cost to each variable. 
\end{enumerate}
The idea is that we keep splitting the Hamiltonian, according to the
variable selected from $C_{H}(x_{i})$, until $C(H)$ is true.

\subsection{Choosing $C(H)$}

Different problems may involve different constraints. If a device
can only handle 2-qubit interactions (such as SQUID-based quantum
annealers as in \cite{Ronnow2014}) we might want a different $C(H)$
than if the device can handle 3-qubit interactions (such as NMR-based
AQCs as in \cite{Xu2012}). If we cannot, or do not want to add any
auxiliary variables, then we do not need the function $C(H)$.

If we wish to allow the addition of auxiliary variables, then for
each term $t$ in $H$, we determine how many auxiliary variables
$n_{{\rm aux},t}$ will be needed in order to reduce $t$ to our desired
order (quadratic order for a device that allows 2-qubit interactions,
cubic order for a device that allows 3-qubit interactions, etc.).
The function is then

\begin{equation}
C(H)=n+\sum_{t}n_{{\rm aux},t}\le Q,\label{eq:cost1}
\end{equation}
where $n$ is the original number of qubits before any auxiliary qubits
were added and $Q$ is our AQC's qubit capacity.

Since the most successful quantum annealing experiments performed
thus far have been on architectures which do not allow higher than
2-qubit interactions, we will give an example of how to choose $C(H)$
for such a device. There are many different ways to quadratize a term
$t$, and each of these methods will have its own $n_{{\rm aux},t}$,
but we know from \cite{Bian2013} that $n_{{\rm aux},t}$ will not
be more than 

\begin{equation}
n_{{\rm aux},t}=\mbox{\ensuremath{\mathcal{R}}}\left({\rm order}(t)-2\right),\label{eq:maximum cost for each term}
\end{equation}
where $\mbox{\ensuremath{\mathcal{R}}}$ is the Ramp function (see
Appendix for details about the quadratization method which only needs
at most this many auxiliary variables). For terms that are already
quadratic, linear, or constant, ${\rm order}(t)\le2$ so $\mbox{\ensuremath{\mathcal{R}}}\left({\rm order}(t)-2\right)=0$
and no auxiliary variables are necessary. If $t$ is, for example,
quintic, then $\mbox{\ensuremath{\mathcal{R}}}\left({\rm order}(t)-2\right)=\mbox{\ensuremath{\mathcal{R}}}\left(5-2\right)=3$
so the \emph{maximum }number of auxiliary qubits added to the cost
function in Eq. \ref{eq:cost1} is 3. 

Therefore, if our goal is to quadratize the Hamiltonian for a device
that only allows up to 2-qubit interactions, and we are limited to
only $Q$ total qubits, then the cost relation is

\begin{equation}
C(H)=n+\sum_{t}\mbox{\ensuremath{\mathcal{R}}}\left({\rm order}(t)-2\right)\le Q.\label{eq:costFunction}
\end{equation}

\subsection{Choosing $C_{H}(x_{i})$}

As in the previous section, our choice of $C_{H}(x_{i})$ depends
on the situation. We may wish to only have quadratic terms without
introducing \emph{any} auxiliary variables, or we may want to choose
a cost function that picks the variable that appears most frequently
in the undesired (super-quadratic) terms. If we choose Eq. \ref{eq:costFunction}
to be our cost formula, we may wish to choose a greedy $C_{H}(x_{i})$
that simply minimizes the number of auxiliary variables that would
need to be added in order to quadratize it. In conjunction with the
cost formula in Eq. \ref{eq:costFunction}, we may define:

\begin{equation}
C_{H}(x_{i})=\sum_{t}\left[I_{x_{i},t}\cdot\mbox{\ensuremath{\mathcal{R}}}\left(\text{order}(t)-2+1\right)\right],\label{eq:varCost}
\end{equation}
where $I_{x_{i},t}$ is $1$ if $x_{i}$ appears in $t$ and 0 otherwise.
The indicator function makes sure we only count terms in which $x_{i}$
appears, and $\mbox{\ensuremath{\mathcal{R}}}\left({\rm order}(t)-2\right)$
is of course the maximum number of auxiliary variables needed to quadratize
term $t$, but we include $+1$ to account for when the variable is
set to $1$. For the splitting, we then choose the variable $x_{i}$
with the biggest $C_{H}(x_{i})$. 

\begin{table*}
\protect\caption{Performance of split-reduc on R(4,3)\label{tab:Split-reduc-on-R(4,3)}}

\begin{centering}
\begin{tabular*}{1\textwidth}{@{\extracolsep{\fill}}cccc}
\hline 
\noalign{\vskip2mm}
Number of vertices & Total size of search space & \# of Hamiltonians needed with $Q=128$ & Upper bound from Section \ref{sub:More-sophisticated-estimates}\tabularnewline[2mm]
\hline 
\noalign{\vskip2mm}
6 & $2^{15}$ & 1 & 1\tabularnewline
7 & $2^{21}$ & 9 & 9\tabularnewline
8 & $2^{28}$ & 169 & 187\tabularnewline
9 & $2^{36}$ & 6~716 & 9~097\tabularnewline[2mm]
\hline 
\noalign{\vskip2mm}
Number of vertices & Total size of search space & \# of Hamiltonians needed with $Q=50$ & Upper bound from Section \ref{sub:More-sophisticated-estimates}\tabularnewline[2mm]
\hline 
\noalign{\vskip2mm}
6 & $2^{15}$ & 9 & 9\tabularnewline[2mm]
7 & $2^{21}$ & 126 & 156\tabularnewline
8 & $2^{28}$ & 3~367 & 3~893\tabularnewline
9 & $2^{36}$ & 177~754 & 346~758\tabularnewline[2mm]
\hline 
\noalign{\vskip2mm}
Number of vertices & Total size of search space & \# of Hamiltonians needed with $Q=30$ & Upper bound from Section \ref{sub:More-sophisticated-estimates}\tabularnewline[2mm]
\hline 
\noalign{\vskip2mm}
6 & $2^{15}$ & 24 & 27\tabularnewline
7 & $2^{21}$ & 398 & 573\tabularnewline
8 & $2^{28}$ & 13~389 & 22~246\tabularnewline
9 & $2^{36}$ & 829~055 & 1~932~743\tabularnewline[2mm]
\hline 
\end{tabular*}
\par\end{centering}

\centering{}\rule[-0.5ex]{1\textwidth}{0.5pt}
\end{table*}

\section{Estimates on the number of splittings}

In Section \ref{sec:A-quick-example} the benefits of splitting were
clear. We only needed 3 objective functions in the end, which is a
small fraction of the search space of size $2^{8}$. But what about
in general? It is not difficult to construct cases in which the number
of splits required blows up. However, this is often not the case.
We may think of the splitting process as giving rise to a binary tree.
The root of the tree is the original objective function and each node
has two branches or zero branches (from splitting or not splitting
respectively). Establishing tight analytic bounds on the number of
leaves may seem tricky, but with simple assumptions, we show that
we can estimate upper and lower bounds remarkably well.

\subsection{Heuristic bounds}

Let us start with the most basic lower bound we can imagine. We can
assume that the shortest path from the root of the tree to a Hamiltonian
that satisfies all hardware requirements is found by successively
choosing the variable with the highest cost and setting it to 0. While
false in cases like $H=(1-x_{1})(1-x_{2})(1-x_{3})$ where setting
any variable to 1 is preferable to setting it to 0, it is usually
true when a lot of the monomials have the same sign or many terms
do not share variables. Likewise, we can assume that the longest path
is found by setting the highest cost variable to 1 at each split. 

Provided the above conditions hold, finding the lengths of the extreme
paths then becomes trivial and requires at most $n$ substitutions.
Once we know these lengths, call them $l$ and $s$ for the longest
and shortest path respectively, we know a lower bound is $2^{s}$
and an upper bound is $2^{l}$.

\subsection{More sophisticated estimates based on combinatorics\label{sub:More-sophisticated-estimates}}

The above bounds are not very tight, so we formulate a more sophisticated
estimate, and we ensure that the method tends to overestimate the
number of splits. Let us make the stronger assumption that if $s$
variables were set to 0 to obtain the shortest path, then setting
$s$ variables to 0 will always be sufficient to obtain a Hamiltonian
that satisfies the hardware requirements (a ``desirable Hamiltonian'').
The reason this tends to overestimate (and hence could be considered
an upper bound) is that it ignores the fact that setting a variable
to 1 also helps simplify the Hamiltonian. Using the same number of
operations as before, we can now find better bounds. We know that
either $s$ variables are set to 0 to obtain a desirable Hamiltonian,
or $l$ variables have been set to 0 or 1 (since $l$ is the length
of the longest path). To count the number of such paths consider an
$l$-bit string

\begin{equation}
x_{1}x_{2}\ldots x_{l}.
\end{equation}

There are ${l \choose s}$ ways to choose at which stage the $s$
variables are set to 0. Filling in all the blank spaces before the
last 0 with 1's and leaving the rest empty characterizes all desirable
Hamiltonians in which $s$ variables were set to 0. If $k$ variables
are set to 0 where $k<s$, then there are ${l \choose k}$ desirable
Hamiltonians since all we need is that $l$ variables have been given
a value. Thus, the number of Hamiltonians is

\begin{equation}
\sum_{k=0}^{s}{l \choose k}.\label{eq:eq1}
\end{equation}

What happens if we do not ignore the reducing potential of setting
a variable to 1? Suppose $R_{i}$ right moves (setting variables to
1) reduces a Hamiltonian as much as $R_{i}$ left moves (setting variables
to 0). We want to count the number of paths where $k$ left moves
are made. Since $k<s$, left moves alone will not simplify our Hamiltonian.
We will need $R_{s-k}$ right moves to make up the difference. However
we can include a full $R_{s-k+1}-1$ right moves since without that
last right move the Hamiltonian will not be desirable. That means
we have $k+R_{s-k+1}-1$ slots to fill with $k$ left moves and $R_{s-k+1}-1$
right moves. The number of such paths is simply ${R_{s-k+1}-1+k \choose k}$
and so the total number of desirable Hamiltonians is

\begin{equation}
1+\sum_{k=1}^{s}{R_{s-k+1}-1+k \choose k}.\label{eq:est}
\end{equation}

This too is likely a slight over-estimate since in practice we often
encounter Hamiltonians that don't require exactly $R_{i}$ right moves,
but rather some number in the neighborhood of $R_{i}$. Now only one
issue remains: calculating $R_{i}$. The authors of this paper prefer
over-estimates to bad estimates, so we shall try to find $R_{i}$
that are likely larger than they need to be. We start by successively
making right moves until a desirable Hamiltonian is reached. Before
each right move, however, we note how many left moves would be needed
to reach a desirable Hamiltonian from this point and thus generate
a sequence of length $l+1$ (the number of nodes on the path). If
the sequence is non-increasing, this method is likely to produce a
good estimate since it conforms to the assumptions we made. We then
define $R_{i}$ to be the position of the last occurrence of $s-i+1$
in the sequence since that is the point at which adding a right move
would remove the need for a left move. If the sequence is not non-increasing
then this will just produce a higher upper bound and if the sequence
skips a number (by for example, decreasing by two), we define $R_{i}$
instead to be the last occurrence of a number larger than $s-i+1$.
This procedure involves $\mathcal{O}(n^{2})$ steps since the max
length of any path is $n$.

To demonstrate the idea, we consider the objective function from Section
\ref{sec:A-quick-example}. The shortest path is $s=1$ and the longest
is $l=2$ (see Fig. 1). The sequence generated by the above procedure
is $(1,1,0)$ so $R_{1}=2$. That means the number of splits is $1+{2-1+1 \choose 1}=3$,
which happens to be correct!

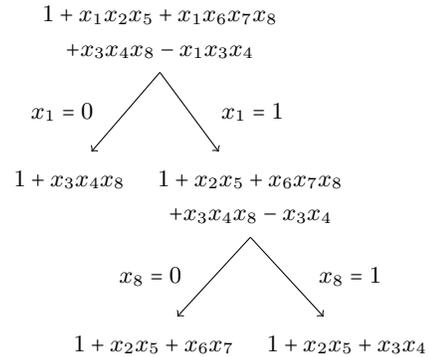
\begin{figure}[h!]
  \centering
  \begin{forest}
  for tree={
    l sep=30pt,
    parent anchor=south,
    align=center,
    edge=->
  }
  [$1+x_{1}x_{2}x_{5}+x_{1}x_{6}x_{7}x_{8}$\\ $+x_{3}x_{4}x_{8}-x_{1}x_{3}x_{4}$
    [$1+x_{3}x_{4}x_{8}$,edge label={node[midway,left]{$x_1=0 \quad$}}
    ]
    [$1+x_{2}x_{5}+x_{6}x_{7}x_{8}$\\ $+x_{3}x_{4}x_{8}-x_{3}x_{4}$,edge label={node[midway,right]{$\quad x_1=1$}}
	  [$1+x_{2}x_{5}+x_{6}x_{7}$,edge label={node[midway,left]{$x_8=0 \quad$}}
      ]
	  [$1+x_{2}x_{5}+x_{3}x_{4}$,edge label={node[midway,right]{$\quad x_8=1$}}
	  ]
    ]
  ]
  \end{forest}
  \caption{A tree representation of the splitting of $f$.}
\end{figure}

\section*{Performance on Ramsey number Hamiltonians}

It has been shown in \cite{Gaitan2012,Bian2013} that finding the
Ramsey number $R(m,n)$ is equivalent to finding what number of vertices
is needed for the ground state of a certain Hamiltonian to have an
energy of greater than 0. For each $(m,n)$ a Hamiltonian is made
to be associated with a graph $G$ with $N$ vertices, and counts
the number of complete subgraphs $K_{m}$, and $n$-independent sets.
The first number $N$ such that the global minimum of $H(m,n,N)$
is not 0, is defined as the Ramsey number $R(m,n)$.

\subsection{$R(m,2)$}

The largest Ramsey number determined by the quantum annealing device
in \cite{Bian2013} was $R(8,2)=8$. The Hamiltonian for this case
is:

\begin{equation}
H=\sum_{k=1}^{L_{m}}1-a_{k}+\prod_{k=1}^{L_{m}}a_{k},\ \ \ \ L_{m}={m \choose 2}=28\label{eq:Ramsey}
\end{equation}
and it is clear we need to deal with a 28 qubit interaction, because
the second term is a product of 28 qubits. In \cite{Bian2013} they
introduce auxiliary variables. We will use split-reduc instead.

Due to the complete symmetry of all the variables we can pick one
at random at each step and split it. If we choose not to allow auxiliary
variables, and aim to split $H$ until it is quadratic, we end up
with the following Hamiltonians after splitting:

\begin{equation}
1+\sum_{k=1+i}^{28}1-a_{k}\text{ for }1\leq i\leq26\text{ and }2-a_{27}-a_{28}+a_{27}a_{28}.
\end{equation}
If we had used Eq. $\ref{eq:est}$ to predict the number of objective
functions we would have found that $s=1,\text{ and }l=R_{1}=26$.
That would mean the number of Hamiltonians is $1+{R_{1} \choose 1}=1+26=27$,
which also happens to be correct! 

For $R(m,2)$ in general the combinatorial estimates in Section \ref{sub:More-sophisticated-estimates}
are provably correct, and $s=1,\ l=R_{1}={m \choose 2}-2$. Thus with
the 128 qubits available in the quantum annealing device of \cite{Bian2013},
the authors could also have calculated $R(16,2)$ with \emph{at most}
${16 \choose 2}-2=118$ runs. Section F of the Supplementary Information
of \cite{Bian2013} explains that the annealing runtimes tend to be
around 2.5~ms, which includes a 1.5~ms delay for reading out the
answer from the machine. Therefore, 118 runs on the device is feasible
within 1~second. It is clear that the quantum annealing device of
\cite{Bian2013} is not so much runtime limited for this problem,
as it is limited by the qubit capacity. We also note that 118 runs
on the quantum annealing device is 35 orders of magnitude smaller
than the size of the total search space if a brute force search were
to be attempted to find $R(16,2)$. 

Furthermore, while $R(16,2)$ was the largest $R(m,2)$ Ramsey number
that could have been determined by the 128 qubit device used in \cite{Bian2013}
had they used split-reduc, we note that the newest version of that
device has a qubit capacity of $Q=2048$, meaning that we could now
determine $R(64,2)$ which requires ${64 \choose 2}=2016$ qubits,
and would only require \emph{at most }${64 \choose 2}-2=2014$ runs
on the device.

\subsection{$R(m,3)$}

The $R(m,2)$ numbers have very simple objective functions. $R(m,3)$
Ramsey numbers are much more complicated, and the only one that was
found in \cite{Bian2013} was $R(3,3).$ The Hamiltonian for $R(4,3)$
for each $N$ is too lengthy to present here, but can be derived from
\cite{Bian2013} and has at most 6-qubit interactions. Therefore,
we apply split-reduc with Eq. $\ref{eq:costFunction}$ as our choice
of $C(H)$ and Eq. $\ref{eq:varCost}$ as our choice of $C_{H}$.
Table \ref{tab:Split-reduc-on-R(4,3)} shows how close our over-estimates
from Section \ref{sub:More-sophisticated-estimates} are for $R(4,3)$,
where we know that the required number of vertices (and hence the
Ramsey number itself) is 9. While minimizing 6716 Hamiltonians would
only take a few seconds on the quantum annealing device of \cite{Bian2013},
we note that this device has \emph{another} restriction, which was
not relevant for $R(m,2)$ because every term after split-reduction
was linear at most (except for the last one). While the split-reduced
$R(4,3)$ Hamiltonians meet the requirement that they are all quadratic
at most, the quantum device of \cite{Bian2013} also requires that
the quadratic couplings can be implemented on their ``chimera''
graph.

In their example, $R(8,2)$ with $N=8$ could be determined with a
Hamiltonian that after quadratization had 54 qubits, and required
30 more qubits to chimerize the connectivity of the 54-vertex graph
describing the connections between all qubits in the Hamiltonian.
Therefore, for $R(4,3)$, if we choose the case in Table \ref{tab:Split-reduc-on-R(4,3)}
that uses at most $Q=50$ qubits in the split-reduced Hamiltonians,
it is reasonable to assume that each of the resulting 177~754 Hamiltonians
could be chimerized using the 72 qubits remaining in the 128-qubit
device. Once again, if each minimization again took 2.5~ms, $R(4,3)$
would be determined within 10 minutes.

\section{Conclusion}

This is the second paper of a 2-part series on techniques for reducing
multi-variable terms in discrete optimization problems. The first
method is called ``deduc-reduc'' because it uses \emph{deductions
}to \emph{reduce }the multi-qubit (multi-variable) terms in the Hamiltonian
(objective function), and is presented in \cite{Tanburn2015a} with
an application to the quantum factorization of numbers larger than
56153, which is currently the largest number factored on a quantum
device \cite{DattaniBryans2014}. Deduc-reduc can also be used to
reduce multi-qubit interactions in the Ramsey number Hamiltonians
discussed in the present paper, but we wished to focus only on the
split-reduc method in this paper. Combining deduc-reduc, split-reduc,
and a third algorithm we have recently devised for reducing the size
of the search space for the Ramsey number discrete optimization problem,
we are able to establish estimated runtimes for some of the presently
undetermined Ramsey numbers such as $R(6,4),$ and $R(10,3)$ \cite{Okada2015a}.

\section*{Appendix: Quadratization method needing at most $\mbox{\ensuremath{\mathcal{R}}}\left({\rm order}(t)-2\right)$
auxiliary qubits}

One way to quadratize a high-order term is to use the penalty function
presented in Section II of the Supplementary Information of \cite{Bian2013}:

\begin{equation}
P(a_{1},a_{2};b)=a_{1}a_{2}-2(a_{1}+a_{2})b+3b,
\end{equation}
which obtains a minimum of $0$ only if $b=a_{1}a_{2}$. Therefore
if our Hamiltonian has a high-order term such as:

\begin{equation}
a_{1}a_{2}a_{3}\ldots a_{n},
\end{equation}
we can reduce its order by one, by replacing $a_{1}a_{2}$ with a
new variable $b$:

\begin{equation}
a_{1}a_{2}a_{3}\ldots a_{n}\rightarrow ba_{3}\ldots a_{n}+\lambda P(a_{1},a_{2};b),\label{eq:transformation for reducing high order terms}
\end{equation}
for a scalar $\lambda$ that is sufficiently large to not introduce
any spurious minima (this is the ``deduc-reduc'' method of Part
1 of this paper \cite{Tanburn2015a}, with $b=a_{1}a_{2}$ as the
deduction, and the choice of $\lambda$ is discussed there). By construction,
whether the LHS or RHS of Eq. \ref{eq:transformation for reducing high order terms}
is considered, the unique minimum/minima will be the same, but the
LHS has order $n$ and the RHS does not have any terms greater than
order $n-1$. 

Our reduced term 
\begin{equation}
ba_{3}\ldots a_{n}
\end{equation}
can then be further reduced by choosing another 2 variables to transform.
Repeatedly applying this method allows us to quadratize a term $t$
with \emph{at most} ${\rm order}(t)-2$ applications, which explains
Eq. $\ref{eq:maximum cost for each term}$ in the main text.

\section*{Acknowledgments}

We gratefully thank Oliver Lunt of Oxford University's Trinity College
for careful proofreading of the manuscript. 

\bibliographystyle{apsrev4-1}

\begin{thebibliography}{11}%
\makeatletter
\providecommand \@ifxundefined [1]{%
 \@ifx{#1\undefined}
}%
\providecommand \@ifnum [1]{%
 \ifnum #1\expandafter \@firstoftwo
 \else \expandafter \@secondoftwo
 \fi
}%
\providecommand \@ifx [1]{%
 \ifx #1\expandafter \@firstoftwo
 \else \expandafter \@secondoftwo
 \fi
}%
\providecommand \natexlab [1]{#1}%
\providecommand \enquote  [1]{``#1''}%
\providecommand \bibnamefont  [1]{#1}%
\providecommand \bibfnamefont [1]{#1}%
\providecommand \citenamefont [1]{#1}%
\providecommand \href@noop [0]{\@secondoftwo}%
\providecommand \href [0]{\begingroup \@sanitize@url \@href}%
\providecommand \@href[1]{\@@startlink{#1}\@@href}%
\providecommand \@@href[1]{\endgroup#1\@@endlink}%
\providecommand \@sanitize@url [0]{\catcode `\\12\catcode `\$12\catcode
  `\&12\catcode `\#12\catcode `\^12\catcode `\_12\catcode `\%12\relax}%
\providecommand \@@startlink[1]{}%
\providecommand \@@endlink[0]{}%
\providecommand \url  [0]{\begingroup\@sanitize@url \@url }%
\providecommand \@url [1]{\endgroup\@href {#1}{\urlprefix }}%
\providecommand \urlprefix  [0]{URL }%
\providecommand \Eprint [0]{\href }%
\providecommand \doibase [0]{http://dx.doi.org/}%
\providecommand \selectlanguage [0]{\@gobble}%
\providecommand \bibinfo  [0]{\@secondoftwo}%
\providecommand \bibfield  [0]{\@secondoftwo}%
\providecommand \translation [1]{[#1]}%
\providecommand \BibitemOpen [0]{}%
\providecommand \bibitemStop [0]{}%
\providecommand \bibitemNoStop [0]{.\EOS\space}%
\providecommand \EOS [0]{\spacefactor3000\relax}%
\providecommand \BibitemShut  [1]{\csname bibitem#1\endcsname}%
\let\auto@bib@innerbib\@empty
\bibitem [{\citenamefont {Tanburn}\ \emph {et~al.}(2015)\citenamefont
  {Tanburn}, \citenamefont {Okada},\ and\ \citenamefont
  {Dattani}}]{Tanburn2015a}%
  \BibitemOpen
  \bibfield  {author} {\bibinfo {author} {\bibfnamefont {R.}~\bibnamefont
  {Tanburn}}, \bibinfo {author} {\bibfnamefont {E.}~\bibnamefont {Okada}}, \
  and\ \bibinfo {author} {\bibfnamefont {N.}~\bibnamefont {Dattani}},\ }\href
  {http://arxiv.org/abs/1508.04816} {\bibfield  {journal} {\bibinfo  {journal}
  {Physical Review A (submitted)}\ } (\bibinfo {year} {2015})},\ \Eprint
  {http://arxiv.org/abs/1508.04816} {arXiv:1508.04816} \BibitemShut {NoStop}%
\bibitem [{\citenamefont {Nemhauser}\ and\ \citenamefont
  {Wolsey}(1981)}]{Nemhauser1981}%
  \BibitemOpen
  \bibfield  {author} {\bibinfo {author} {\bibfnamefont {G.}~\bibnamefont
  {Nemhauser}}\ and\ \bibinfo {author} {\bibfnamefont {L.}~\bibnamefont
  {Wolsey}},\ }\href {\doibase 10.1016/S0304-0208(08)73471-6} {\emph {\bibinfo
  {title} {North-Holland Mathematics Studies}}},\ \bibinfo {series}
  {North-Holland Mathematics Studies}, Vol.~\bibinfo {volume} {59}\ (\bibinfo
  {publisher} {Elsevier},\ \bibinfo {year} {1981})\ pp.\ \bibinfo {pages}
  {279--301}\BibitemShut {NoStop}%
\bibitem [{\citenamefont {Orlin}(2007)}]{Orlin2007}%
  \BibitemOpen
  \bibfield  {author} {\bibinfo {author} {\bibfnamefont {J.~B.}\ \bibnamefont
  {Orlin}},\ }\href {\doibase 10.1007/s10107-007-0189-2} {\bibfield  {journal}
  {\bibinfo  {journal} {Mathematical Programming}\ }\textbf {\bibinfo {volume}
  {118}},\ \bibinfo {pages} {237} (\bibinfo {year} {2007})}\BibitemShut
  {NoStop}%
\bibitem [{\citenamefont {Jegelka}\ \emph {et~al.}(2011)\citenamefont
  {Jegelka}, \citenamefont {Lin},\ and\ \citenamefont {Bilmes}}]{Jegelka2011}%
  \BibitemOpen
  \bibfield  {author} {\bibinfo {author} {\bibfnamefont {S.}~\bibnamefont
  {Jegelka}}, \bibinfo {author} {\bibfnamefont {H.}~\bibnamefont {Lin}}, \ and\
  \bibinfo {author} {\bibfnamefont {J.~A.}\ \bibnamefont {Bilmes}},\ }in\ \href
  {http://papers.nips.cc/paper/4348-on-fast-approximate-submodular-minimization}
  {\emph {\bibinfo {booktitle} {Advances in Neural Information Processing
  Systems}}}\ (\bibinfo {year} {2011})\ pp.\ \bibinfo {pages}
  {460--468}\BibitemShut {NoStop}%
\bibitem [{\citenamefont {Boros}\ and\ \citenamefont
  {Hammer}(2002)}]{Boros2002}%
  \BibitemOpen
  \bibfield  {author} {\bibinfo {author} {\bibfnamefont {E.}~\bibnamefont
  {Boros}}\ and\ \bibinfo {author} {\bibfnamefont {P.~L.}\ \bibnamefont
  {Hammer}},\ }\href {\doibase 10.1016/S0166-218X(01)00341-9} {\bibfield
  {journal} {\bibinfo  {journal} {Discrete Applied Mathematics}\ }\textbf
  {\bibinfo {volume} {123}},\ \bibinfo {pages} {155} (\bibinfo {year}
  {2002})}\BibitemShut {NoStop}%
\bibitem [{\citenamefont {Bian}\ \emph {et~al.}(2013)\citenamefont {Bian},
  \citenamefont {Chudak}, \citenamefont {Macready}, \citenamefont {Clark},\
  and\ \citenamefont {Gaitan}}]{Bian2013}%
  \BibitemOpen
  \bibfield  {author} {\bibinfo {author} {\bibfnamefont {Z.}~\bibnamefont
  {Bian}}, \bibinfo {author} {\bibfnamefont {F.}~\bibnamefont {Chudak}},
  \bibinfo {author} {\bibfnamefont {W.~G.}\ \bibnamefont {Macready}}, \bibinfo
  {author} {\bibfnamefont {L.}~\bibnamefont {Clark}}, \ and\ \bibinfo {author}
  {\bibfnamefont {F.}~\bibnamefont {Gaitan}},\ }\href {\doibase
  10.1103/PhysRevLett.111.130505} {\bibfield  {journal} {\bibinfo  {journal}
  {Physical Review Letters}\ }\textbf {\bibinfo {volume} {111}},\ \bibinfo
  {pages} {130505} (\bibinfo {year} {2013})}\BibitemShut {NoStop}%
\bibitem [{\citenamefont {Ronnow}\ \emph {et~al.}(2014)\citenamefont {Ronnow},
  \citenamefont {Wang}, \citenamefont {Job}, \citenamefont {Boixo},
  \citenamefont {Isakov}, \citenamefont {Wecker}, \citenamefont {Martinis},
  \citenamefont {Lidar},\ and\ \citenamefont {Troyer}}]{Ronnow2014}%
  \BibitemOpen
  \bibfield  {author} {\bibinfo {author} {\bibfnamefont {T.~F.}\ \bibnamefont
  {Ronnow}}, \bibinfo {author} {\bibfnamefont {Z.}~\bibnamefont {Wang}},
  \bibinfo {author} {\bibfnamefont {J.}~\bibnamefont {Job}}, \bibinfo {author}
  {\bibfnamefont {S.}~\bibnamefont {Boixo}}, \bibinfo {author} {\bibfnamefont
  {S.~V.}\ \bibnamefont {Isakov}}, \bibinfo {author} {\bibfnamefont
  {D.}~\bibnamefont {Wecker}}, \bibinfo {author} {\bibfnamefont {J.~M.}\
  \bibnamefont {Martinis}}, \bibinfo {author} {\bibfnamefont {D.~A.}\
  \bibnamefont {Lidar}}, \ and\ \bibinfo {author} {\bibfnamefont
  {M.}~\bibnamefont {Troyer}},\ }\href {\doibase 10.1126/science.1252319}
  {\bibfield  {journal} {\bibinfo  {journal} {Science}\ }\textbf {\bibinfo
  {volume} {345}},\ \bibinfo {pages} {420} (\bibinfo {year}
  {2014})}\BibitemShut {NoStop}%
\bibitem [{\citenamefont {Xu}\ \emph {et~al.}(2012)\citenamefont {Xu},
  \citenamefont {Zhu}, \citenamefont {Lu}, \citenamefont {Zhou}, \citenamefont
  {Peng},\ and\ \citenamefont {Du}}]{Xu2012}%
  \BibitemOpen
  \bibfield  {author} {\bibinfo {author} {\bibfnamefont {N.}~\bibnamefont
  {Xu}}, \bibinfo {author} {\bibfnamefont {J.}~\bibnamefont {Zhu}}, \bibinfo
  {author} {\bibfnamefont {D.}~\bibnamefont {Lu}}, \bibinfo {author}
  {\bibfnamefont {X.}~\bibnamefont {Zhou}}, \bibinfo {author} {\bibfnamefont
  {X.}~\bibnamefont {Peng}}, \ and\ \bibinfo {author} {\bibfnamefont
  {J.}~\bibnamefont {Du}},\ }\href {\doibase 10.1103/PhysRevLett.108.130501}
  {\bibfield  {journal} {\bibinfo  {journal} {Physical Review Letters}\
  }\textbf {\bibinfo {volume} {108}},\ \bibinfo {pages} {130501} (\bibinfo
  {year} {2012})}\BibitemShut {NoStop}%
\bibitem [{\citenamefont {Gaitan}\ and\ \citenamefont
  {Clark}(2012)}]{Gaitan2012}%
  \BibitemOpen
  \bibfield  {author} {\bibinfo {author} {\bibfnamefont {F.}~\bibnamefont
  {Gaitan}}\ and\ \bibinfo {author} {\bibfnamefont {L.}~\bibnamefont {Clark}},\
  }\href {\doibase 10.1103/PhysRevLett.108.010501} {\bibfield  {journal}
  {\bibinfo  {journal} {Physical Review Letters}\ }\textbf {\bibinfo {volume}
  {108}},\ \bibinfo {pages} {010501} (\bibinfo {year} {2012})}\BibitemShut
  {NoStop}%
\bibitem [{\citenamefont {Dattani}\ and\ \citenamefont
  {Bryans}(2015)}]{DattaniBryans2014}%
  \BibitemOpen
  \bibfield  {author} {\bibinfo {author} {\bibfnamefont {N.~S.}\ \bibnamefont
  {Dattani}}\ and\ \bibinfo {author} {\bibfnamefont {N.}~\bibnamefont
  {Bryans}},\ }\href {http://arxiv.org/abs/1411.6758} {\bibfield  {journal}
  {\bibinfo  {journal} {Physical Review Letters (revisions requested)}\ }
  (\bibinfo {year} {2015})},\ \Eprint {http://arxiv.org/abs/1411.6758}
  {arXiv:1411.6758} \BibitemShut {NoStop}%
\bibitem [{\citenamefont {Okada}\ \emph {et~al.}(2015)\citenamefont {Okada},
  \citenamefont {Tanburn},\ and\ \citenamefont {Dattani}}]{Okada2015a}%
  \BibitemOpen
  \bibfield  {author} {\bibinfo {author} {\bibfnamefont {E.}~\bibnamefont
  {Okada}}, \bibinfo {author} {\bibfnamefont {R.}~\bibnamefont {Tanburn}}, \
  and\ \bibinfo {author} {\bibfnamefont {N.~S.}\ \bibnamefont {Dattani}},\
  }\href@noop {} {\bibfield  {journal} {\bibinfo  {journal} {Physical Review A
  (in preparation)}\ } (\bibinfo {year} {2015})}\BibitemShut {NoStop}%
\end{thebibliography}

\end{document}